\documentclass[twocolumn,preprintnumbers,amsmath,amssymb]{revtex4}
\input{epsf}

\newcommand{\be}{\begin{equation}}
\newcommand{\ee}{\end{equation}}
\newcommand{\bea}{\begin{eqnarray}}
\newcommand{\eea}{\end{eqnarray}}
\newcommand{\bef}{\begin{figure}}
\newcommand{\eef}{\end{figure}}

\newcommand{\simge}{\,{}^>_{\sim}\,}
\newcommand{\simle}{\,{}^<_{\sim}\,}

\def\H{\rm H}
\def\h#1{$^{#1}$H}
\def\he#1{$^{#1}$He}
\def\li#1{$^{#1}$Li}
\def\be#1{$^{#1}$Be}
\def\Ob{$\Omega_b$}
\def\Obh{$\Omega_bh^2$}

\def\eps@scaling{0.96}

\def\showone#1{
  \centering
  \leavevmode
  \epsfxsize=\eps@scaling\linewidth
  \epsfbox{#1.eps}
}

\def\epstwo@scaling{0.48}

\def\showtwo#1#2{
  \centering
  \leavevmode
  \epsfxsize=\epstwo@scaling\linewidth
  \epsfbox{#1.eps} \hfil
  \epsfxsize=\epstwo@scaling\linewidth
  \epsfbox{#2.eps}
}

\begin{document}

\title{Did Something Decay, Evaporate, or Annihilate during Big Bang Nucleosynthesis?}
\author{Karsten Jedamzik} 
\affiliation{Laboratoire de Physique Math\'emathique et Th\'eorique, C.N.R.S.,
Universit\'e de Montpellier II, 34095 Montpellier Cedex 5, France}

\begin{abstract}
Results of a detailed examination of the cascade nucleosynthesis resulting from the
putative hadronic decay, evaporation, or annihilation of a primordial relic
during the Big Bang nucleosynthesis (BBN) era are presented. It is found that
injection of energetic nucleons around cosmic time $10^3$sec may lead to an
observationally favored reduction of the primordial \li7/H yield by a 
factor $2 - 3$. Moreover, such sources also generically predict the production
of the \li6 isotope with magnitude close to the as yet unexplained high
\li6 abundances in low-metallicity stars. The simplest of these models
operate at fractional contribution to the baryon density \Obh$\simge 0.025$,
slightly larger than that inferred from standard BBN. Though further 
study is required, such sources, as for example due to
the decay of the next-to-lightest supersymmetric particle into GeV
gravitinos
or the decay of an unstable 
gravitino in the TeV range of abundance 
$\Omega_{\tilde{G}}h^2\sim 5\times 10^{-4}$ show promise to explain both
the \li6 and \li7 abundances in low metallicity stars.
\end{abstract}


\maketitle

Big Bang nucleosynthesis has since long been known as one of the most precise 
and furthest back-reaching probes of cosmic conditions and the cosmic
matter content of the early Universe. It thus, for example, has significantly
contributed to the notion that a large fraction of matter in present-day galaxies
is believed to be of non-baryonic nature as well as considerably limited some
extensions of the standard model of particle physics. Paramount to having become
such a useful tool of cosmology was and is, not only the examination of early
synthesis of light elements in it's simplest standard version, but also
in a variety of non-standard, alternative scenarios, relaxing
a priori non-verified assumptions entering the calculations 
of a standard BBN (SBBN). Depending on the light element yields obtained in these 
latter scenarios, such calculations may then either favor a modified version of BBN, or
strengthen the case for the SBBN. In either case, calculations of
non-standard BBN may be used to place limits on the cosmic condition in the early
Universe. In this spirit, technically advanced calculations of BBN including decaying
particles during, or after BBN, an inhomogeneous baryon distribution, 
small-scale antimatter domains, neutrino degeneracy, or varying fundamental constants
(for reviews cf.~\cite{reviews}), 
among others, have been performed over the years, rendering
SBBN (also by the principle of Occam's razor) as our preferred scenario for BBN

On the observational side significant advances have been made with the advent of 
high-resolution spectrographs
on the Keck- and VLT-telescopes and the resulting capability to perform
D/H abundance determinations of unprecedented accuracy in a few simple 
high-redshift quasar absorption line (QAL) systems. Furthermore an independent 
determination of the fractional contribution of baryons to
the critical density, \Ob, from precision measurements of the cosmic microwave
background radiation (CMB) by various balloon missions and the 
WMAP~\cite{Sper:03} sattelite
has become feasible.
These observations together with the
ever-continuing observational and theoretical efforts to deduce
precise primordial \he4/H- and \li7/\H- ratios, leave BBN in an essentially
observationally overconstrained state, opening the possibility to question 
internal
consistency of the predictions of SBBN. If such a check 
is performed it shows that SBBN is
to first approximation internally consistent, 
though inconsistencies or tensions between
predicted- and observationally inferred- abundances may exist at higher order.
It is the subject of this paper to propose a scenario of non-standard BBN
which may remove some of these tensions.

If the central value of the observationally determined 
D/H$=2.78^{+0.44}_{-0.38}\times 10^{-5}$~\cite{Kirk:03,remark0} by the
average of five QAL systems is taken, SBBN predicts \li7/\H $\approx 4.16\times 10^{-10}$, 
a \he4 mass fraction
$Y_p \approx 0.2480$ at an \Obh$\,\approx 0.0218$.
I refrain from the common practice to give error bars due to
nuclear reaction rate uncertainties on the
theoretical predictions, as there exist surprisingly large differences
between the central values obtained by 
different groups, often larger than the quoted
error bars, particularly in the case of \li7~\cite{remark1}. 
The predicted \li7 should be compared to the observationally
inferred primordial \li7/\H $= 1.23^{+0.34}_{-0.16}\times 10^{-10}$~\cite{Ryan:99}
abundance from the inferred \li7 in atmospheres 
of extreme Pop II stars belonging to the Spite plateau (where the analysis
corrects for \li7 production by cosmic rays), or to the inferred 
\li7 $= 2.19^{+0.30}_{-0.27}\times 10^{-10}$~\cite{Boni:02}
from low-metallicity stars within the globular cluster NGC 6397, indicating
if taken at face value, that there may be a problem in SBBN. This discrepancy
may not be resolved by nuclear reaction rate uncertainties in the main
lithium-producing reaction \he3$(\alpha ,\gamma )$\be7~\cite{Cybu:03} and
only very unlikely due to uncertainties in the lithium-destroying reaction
\be7$(d,p)2$\he4~\cite{Coc:04}. It is conceivable, however, that it is 
due to other
systematic uncertainties entering the inference of primordial 
\li7 abundances, such as stellar astration of \li7 in 
low-metallicity stars, 
or imprecise determinations of stellar surface temperatures in these stars.
Recent 3D non-LTE (i.e. local-thermodynamic-equilibrium) calculations~\cite{Aspl:03} of
lithium lines in low-metallicity stars indicate, however, 
that simplifying assumptions
concerning the stellar atmosphere probably
do not introduce excessively large systematic errors. 

Stellar \li7 may be destroyed by \li7$(p,\alpha )$\he4 
when transported deep enough into the interior of the star, as for example 
due to stellar rotation induced turbulence or 
diffusion. The mixing possibilities
proposed are numerous, however, many fail on the requirement
to deplete \li7 in different stars of different surface temperatures and
rotation velocities by a substantial factor 
($\sim 2-4\equiv 0.3 - 0.6$ dex) neither 
introducing scatter in the observationally inferred \li7/H  nor a 
trend of depletion with surface temperature. Moreover, \li7 has to
be depleted without substantial depletion of the more fragile 
\li6~\cite{Lemo:97,Vang:99} isotope 
observed in at least two of these stars along with that of
\li7. The
large abundance of the \li6 isotope at such low metallicity is anyway
theoretically challenging, even in the absence of stellar depletion 
(see below).
Recent detailed stellar studies~\cite{Sala:01,Thea:01} and
an analysis~\cite{Pins:01} 
of the samples of \li7/H abundances employed to infer
the primordial \li7/H ratio claim that, under certain conditions,
a uniform depletion of 0.2-0.3 dex may be possible, though there
remains lack of observational indication for this, in fact, to be
the case~\cite{Boni:02}.  

Concerning the abundance of \he4 there exists as well a mismatch
between the relatively high prediction and the lower
observationally inferred value 
$Y_p = 0.02390\pm 0.0020$~\cite{Luri:03} (with 
another group finding $Y_p = 0.2421\pm 0.0021$~\cite{Izot:03}). 
I will not much further
elaborate on this discrepancy here, since the abundance of \he4, 
whose observational determination is plagued by various
systematic uncertainties, will hardly change in the scenarios I consider.
Last but not least, there is the vital comparison between the \Obh as
inferred from the CMB and \Obh as preferred by SBBN and D/H.
The inferred baryon density when the WMAP data only is 
taken~\cite{Sper:03} (with a 
power-law $\Lambda$CDM model) is \Obh$ = 0.024\pm 0.001$, whereas
a running spectral index $\Lambda$CDM model results in a value lower
by 0.001. When data from other experiments, of CMB and/or
large-scale structure, probing smaller scales than WMAP are included,
the estimated \Obh drops to around 0.0225 (for combined likelihoods
between \Obh estimates from CMB and SBBN cf. to Ref.~\cite{Cuoc:03}).  
In general the agreement between CMB and SBBN (plus D/H) is excellent,
with nevertheless CMB inferred \Obh values typically staying on the
high side of those preferred by SBBN. Thus non-standard BBN scenarios
are not allowed to operate with \Obh much different than those
of SBBN. 
 
At such large \Obh the bulk of \li7 is produced as \be7
(which is after BBN converted to \li7). It is a not
widely recognized effect that \be7 may be prematurely destroyed via
the reaction chain \be7$(n,p)$\li7 and \li7$(p,\alpha )$\he4 by
significant factors when towards the end of BBN, at
approximate temperatures $30-50\,$keV a small excess of free
neutrons ${\rm n/p}\sim 10^{-5}$ 
as compared to SBBN exists. This phenomena has been observed in both
studies of inhomogeneous BBN~\cite{Jeda:94} 
as well as studies with hadronically
decaying particles~\cite{Reno:88}. 
Such an excess could thus possibly result from
the hadronic decay or evaporation of a relic of the early 
Universe. Thermal neutrons injected at these temperatures are predominantly
incorporated into D via p$(n,\gamma )$D, with a small 
fraction $\sim 10^{-4} - 10^{-5}$ causing the premature
conversion of \be7 to \li7. It is straightforward to 
show that to convert one \be7 nucleus 
$\langle\sigma v\rangle_{\rm p\, n}/\langle\sigma 
v\rangle_{\rm ^7Be\, n}\times {\rm H/^7Be}$
neutrons are required. Due to the \be7$(n,p)$\li7 reaction being between
mirror nuclei, the rate ratio in the above is very small 
$\approx 1.4\times 10^{-5}$.
This implies that the conversion of essentially 
all \be7 to \li7 is accompanied by only a mild 
(\be7/H -independent) excess of D/H $\approx 1.4\times 10^{-5}$,
when compared to SBBN. Of course, not all of the thus produced \li7 will
subsequently be destroyed via \li7$(p,\alpha )$\he4, 
in particular at low temperatures where the
Coulomb barrier is preventing the reaction. Neutrons which are injected
at slightly higher temperatures are further processed into \he4.
Nevertheless, this possible excess in \he4 is essentially negligible due
to the small numbers of neutrons required to destroy \be7.

I have slightly modified the Kawano code in order to test for this
effect. In particular, I have injected thermal neutrons with rate
$dn_n/dt \sim {\rm exp}(-t/\tau)/\tau$ and decay time $\tau = 700\,$s
employing a total injected neutron abundance corresponding
to $\Omega_nh^2\approx 10^{-6}$. In Fig. 1 the light-element synthesis
in the presence (and absence) of such a neutron source is shown, 
assuming $\Omega_bh^2 = 0.026$. It is seen that the neutrons
in fact yield the desired destruction of \be7 and some comparatively
smaller enhancement in \li7. The yields for D/H and \li7/H, with (and without)
extra neutrons are $3.25\times 10^{-5}$, ($2.09\times 10^{-5}$), and
$1.73\times 10^{-10}$, ($5.92\times 10^{-10}$), respectively,
resulting in a 0.53 dex "depletion" of the \li7 yield. The
\he4 abundance remains virtually unchanged (except for an increase
by $\Delta Y_p \approx 0.002$ on account of the increased \Obh with
respect to the above quoted value).

\bef
\epsfxsize=8.5cm
\epsffile[85 50 410 302]{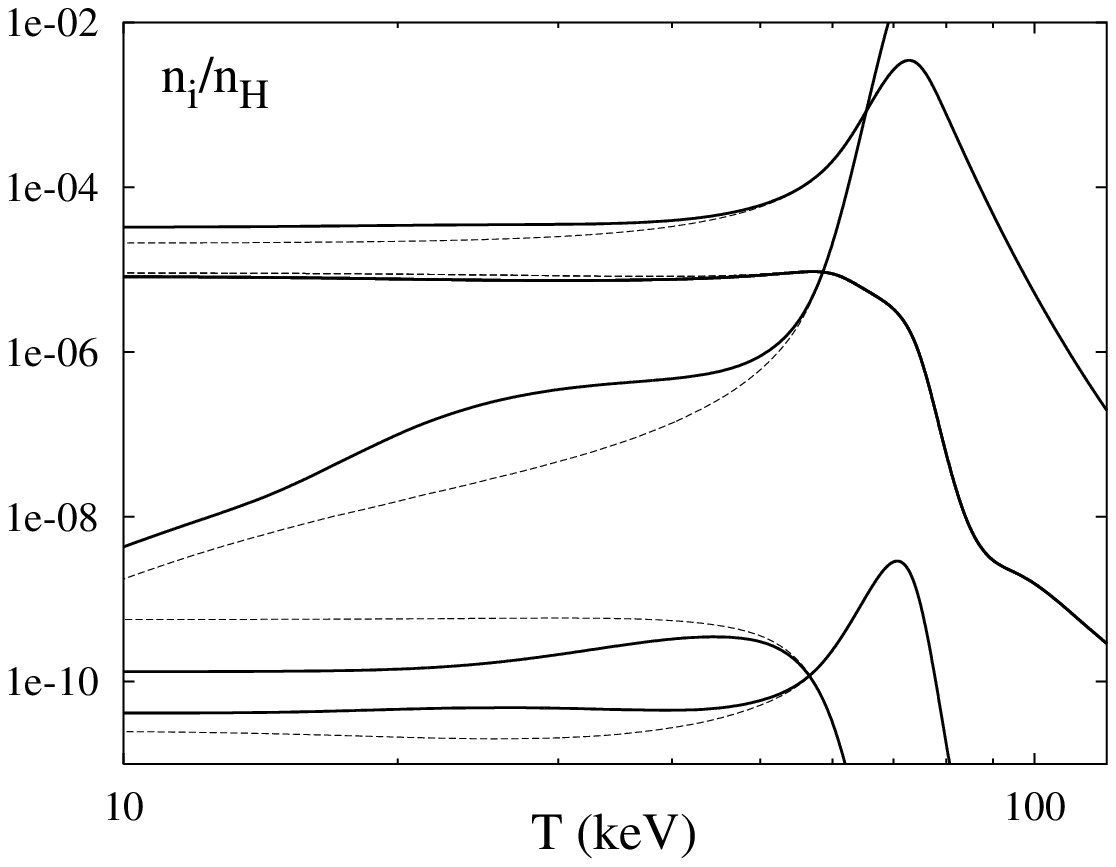}
\caption{Light element evolution as function of temperature with
(solid) and without (dashed) a thermal neutron source. Shown are,
from top to bottom (at lower temperatures), the number ratios 
$n_{\rm D}/n_p$, $n_{\rm ^3He}/n_p$, $n_n/n_p$, $n_{\rm ^7Be}/n_p$,
and $n_{\rm ^7Li}/n_p$, respectively. See text for further detail.}
\label{fig1}
\eef

The effect is encouraging and may present a possible resolution to
the \li7 problem in SBBN. Nevertheless, many viable hadronically
decaying, evaporating, or annihilating candidates inject energetic
hadrons rather than
thermal neutrons. A massive $m_{\tilde{G}}\sim 200\,$GeV 
decaying particle, for example, would lead to the injection of protons
and neutrons with typical energies $\sim 8\,$GeV reaching up to several
tens of GeVs. These are accompanied, of course, by much larger numbers
of pions, $e^{\pm}$, neutrinos, and photons. Cascade nucleosynthesis
due to energetic electromagnetically interacting particles may not
be of immediate interest here, not only because it is operative at lower 
$T\simle 3\,$ keV, but also since the small required neutron densities
to resolve the \li7 discrepancy
may not imply much of an effect. 
Similarly, n$\leftrightarrow$p interconversion by 
mesons~\cite{Reno:88,Kohr:01}, 
potentially important
at higher temperatures, has a negligible effect due to the smallness
of the assumed perturbation. 
I caution though that these conclusions are dependent
on the injected meson- and baryon- multiplicities, and the ratio of 
electromagnetically to hadronically interacting particles, and are strictly
only valid in the context of strongly interacting jet dynamics.
In contrast, spallation of \he4 by energetic nucleons is important, and
has heretofore only been considered after BBN by the 
pioneering study of Ref.~\cite{Dimo:88} (cf. also to~\cite{Khlop}), 
and with an amount of injected nucleons far larger than of interest here.
Similar holds for 
nonthermal nucleosynthesis, in particular, the reactions
\h3$(\alpha ,n)$\li6 and \he3$(\alpha ,p)$\li6 with energy threshold
induced by energetic
\h3 and \he3 (themselves produced by the spallation of \he4) are of
paramount importance for the \li6 abundance. 


The large abundance of \li6 in low-metallicity Pop II stars is a bit of a 
mystery. It has been observed in at least three stars at low
metallicity [Fe/H]$< -2$ with the best case probably the star
HD84937 with \li6/\li7$= 0.052\pm 0.019$~\cite{li6:highZ},
as well as in two stars at higher metallicity 
[Fe/H]$\approx  -0.6$~\cite{li6:lowZ}, with all abundance ratios
coincidentally being similar~\cite{remark2}. \li6 is traditionally
not thought to be of primordial origin due to the cross section
D$(\alpha ,\gamma )$\li6, absent of threshold, being 
small~\cite{Noll:97}. \li6 is thus believed to have its origin in
cosmic-ray nucleosynthesis, being produced along-side with
\li7 (to be added to the BBN yield), 
\be9, $^{10}$B, and $^{11}$B via spallation 
(${\rm p}, \alpha + {\rm CNO} \to {\rm LiBeB}$)
and fusion ($\alpha + \alpha \to {\rm Li}$) reactions by galactic
cosmic rays. It is rather controversial as to 
whether~\cite{pro:li6}, or not~\cite{Rama:99,Alib:02}, it is possible to 
produce the large observed \li6 abundance in Pop II     
stars via cosmic rays generated by thermonuclear or core-collapse
supernovae, even when so far unknown cosmic ray populations
(e.g., low-energy and metallicity-dependent) 
are postulated. The problem, in general, seems to be that in order to
reproduce the approximate linear relationship between \be9/H and [Fe]
metal-enriched cosmic rays are strongly favored. Such metal-enriched
cosmic rays yield typical spallation \li6/\be9-ratios $\sim 5-20$ 
agreeing with the ratio $\approx 6$
in the solar system, but not with the ratio
$\sim 80$ as observed in Pop II stars. It may be that there
exists an enhanced fusion contribution (with 
cosmic rays at $\sim$ 30 MeV/nucleon)
at high redshifts, and though it may be energetically problematic 
when associated with stellar evolution~\cite{Rama:99}, it could be due to
virialization shocks during structure formation~\cite{Suzu:02}.
Nevertheless, the origin of \li6 in Pop II stars seems currently
controversial.
This situation has led me to consider alternative production
of \li6 due to non-thermal nucleosynthesis 
after BBN~\cite{Jeda:00, remark3}, resulting from electromagnetic
cascade nucleosynthesis.

Production of \li6 due to nuclear spallation and fusion reactions
towards the end of BBN may as well be very efficient. 
I have performed a detailed Monte-Carlo analysis of the cascade 
nucleosynthesis resulting from 
the decay, evaporation, or annihilation of strongly interacting particles
or defects. To reach a precision to allow for a meaningful
comparison to the comparatively accurate observational determinations
of D/H, \li6/H, \li7/H, and \Obh, care had to be taken.
The analysis includes a detailed modeling of elastic
nucleon-nucleon and nucleon-\he4 scattering. 
It includes also their important inelastic ($pp, np\to pp,pn,nn +\pi 's$ and
$p,n{\rm ^4He\to {}^3He,{}^3H, D}'s + n's,p's +\pi's$) counterparts employing
detailed cross section data. A careful treatment of the recoil energies
of \he4 (from elastic scattering), as well as \h3 and \he3 
(from \he4 spallation)
has been incorporated.
These recoil energy distribution functions are important for a
determination of \li6 yields resulting from the reactions
\h3$(\alpha ,n)$\li6, \he3$(\alpha ,p)$\li6, \he4$(\alpha ,2p)$\he6, 
\he4$(\alpha ,pn)$\li6. Finally, as thermalization of nucleons,
and nuclei, is
a competition between nuclear scattering and Coulomb interactions with
the plasma, in the case of protons and nuclei, and nuclear scattering
and magnetic moment scattering on $e^{\pm}$, in the case of neutrons,
these latter interactions also had to be included~\cite{remark4}.
A detailed account of the analysis, which is beyond the scope
of the present paper, will be presented elsewhere.
These processes have been coupled to the Kawano code~\cite{Smit:93} 
(updated by the NACRE~\cite{Angu:99} reaction compilation).
Finally, the primary spectrum of injected nucleons is computed
with the help of the jet fragmentation code PYTHIA~\cite{pythia}

\bef
\epsfxsize=8cm
\epsffile[100 75 300 415]{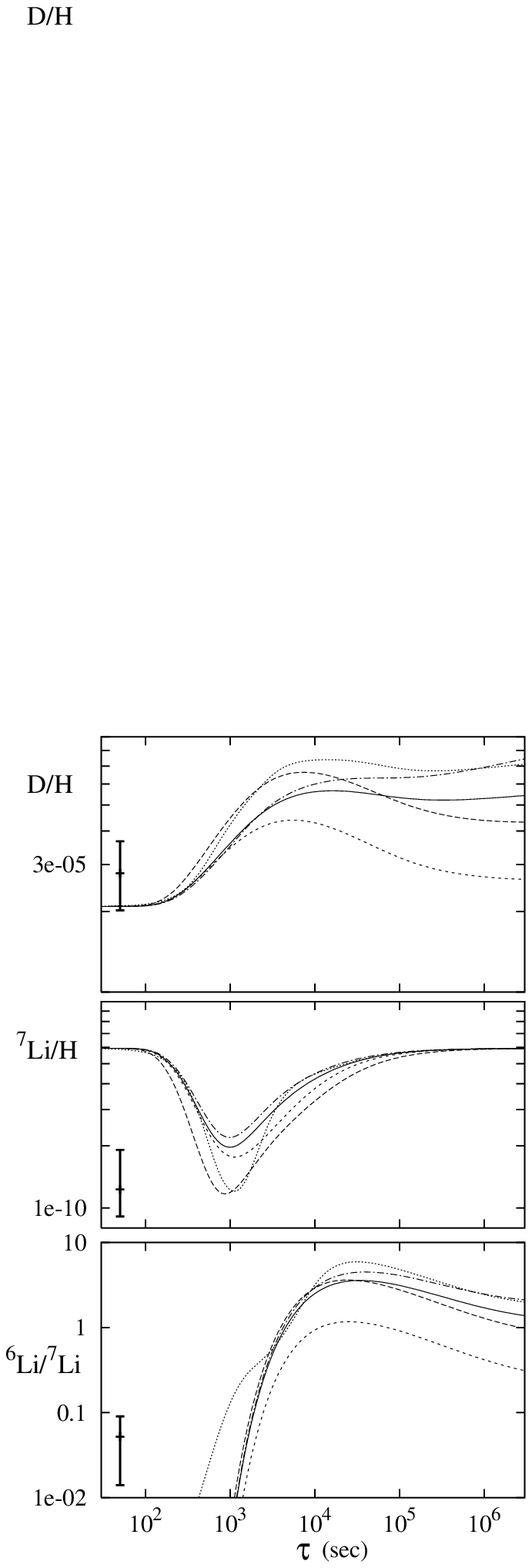}
\caption{Abundance yields of D/H, \li7/H, and \li7/\li6 in an
$\Omega_bh^2 = 0.026$ Universe as
function of the hadronic decay time $\tau$ of a putative 
primordial relic. The models are decay of a $m_{\chi} = 10\,$GeV particle (long-dashed),
decay of a $m_{\chi} = 200\,$GeV particle (solid), decay of a 
$m_{\chi} = 4\,$TeV particle (dashed-dotted), 
injection of monoenergetic nucleons of 
$E_{kin} = 250\,$MeV (short-dashed),
and extended power-law injection due to a 
$m_{\chi} = 200\,$GeV particle (dotted). Also shown are the
two-sigma ranges of the inferred primordial D/H and \li7/H 
abundances~\cite{Kirk:03,Ryan:99} as well as the \li6/\li7 ratio as
inferred in the low-metallicity star HD84937~\cite{li6:highZ}.
See text for further
details.}
\label{fig2}
\eef

The development of a nuclear cascade may be 
summarized as follows. For temperatures above $T\simge 20\,$keV
protons are essentially exclusively thermalized by electromagnetic
interactions with the still abundant $e^{\pm}$ pairs. At lower 
temperatures, thermalization for energetic protons (and nuclei)
$E_{\rm kin}\simge 1\,$GeV
occurs via scattering on plasma protons and \he4, as well as
Thomson scattering on CMBR photons, while less energetic
protons loose energy via Coulomb scattering and plasmon excitation.
In contrast, neutrons thermalize on plasma protons and \he4 up to
temperatures of $T\approx 50-60\,$keV. For higher $T$ the dominant energy
loss is due to $ne^{\pm}$-scattering, with the most rapid fractional
energy loss for energetic neutrons ($\gamma\gg 1$).
Thermalization of nucleons and nuclei
occurs rapidly when compared to the cosmic expansion rate.
Interactions of rapid nucleons on protons (and \he4) at $E_{\rm kin}\simge 1\,$GeV
are to a large fraction ($\sim 0.75$) inelastic,
accompanied by the production of pions.
During such scatterings an interconversion of protons to neutrons
occurs frequently, such that energetic protons produce secondary neutrons. 
For example, though the decay of
a $200\,$GeV particle generates only about $\approx 1$ neutron 
per annihilation,
around $\approx 1, 0.6$ secondary neutrons result at $T \approx 20, 40\,$keV,
respectively~\cite{remark7}, and $\approx 3.5$ asymptotically at low
temperatures $T\sim 0.1-1\,$keV. 
Here at higher temperatures the number of secondary neutrons reduces
due to the rapid Coulomb losses of protons. 
Neutrons, on the other hand, do not
possess a significant bias towards producing secondary neutrons in $np$ inelastic
interactions. Excess neutrons at $T\approx 40\,$keV are mostly due to
inelastic processes on \he4, accompanied by the production of D and \he3
(i.e. $n + {\rm ^4He} \to {\rm D} + p + 2n$, ...),
with a comparatively smaller amount of neutrons removed in pionic fusion processes
(i.e. $np\to {\rm D}\pi^0$, ...). One thus obtains approximately a ratio
n/D$\,\approx 3.6$ for a $200\,$GeV particle at $T\approx 40\,$keV, with
similar ratios for n/\h3 and n/\he3. As the \h3 and \he3 are energetic they
may yield the production of \li6. Nevertheless, \li6 production (and survival)
may only be efficient at somewhat lower temperatures. Due to Coulomb losses
of energetic \h3 and \he3 production is only efficient at $T\simle 20\,$keV,
whereas survival
of the freshly synthesized \li6 against destruction
via \li6$(p,\alpha)$\he3 is only nearly complete for $T\simle 10\,$keV. 
The production of \li6 at temperatures $T\approx 10-20\,$keV
for a $200\,$GeV particle 
is found to be approximately $2\times 10^{-4}$ per decaying particle,
becoming significantly lower at lower temperatures (e.g. $3\times 10^{-5}$ at
$T\approx 1\,$keV).     
Cascade yields are subject to some nuclear physics data uncertainties
which in the case of \li6 may be of the order of a factor two.
In particular, it may be that \li6 yields are underestimated due to
an experimentally incomplete determination of the 
high-energy tail of the energy distribution
of energetic \h3 and \he3 produced in \he4 spallation. 

The developed code allows me to present detailed predictions
on the BBN in the presence of decaying particles.
Figure 2 shows the light-element yields for a variety of decaying particles
as a function of particle life time $\tau$. The panels show, from top-to-bottom,
final abundances of D/H, \li7/H, and \li6/\li7, with the understanding that
$Y_p$ is virtually unchanged when compared to SBBN at the same \Obh.
In all models \Obh$=0.026$ has been assumed. Hadronically decaying particle
yields (with the simplifying assumption that $\chi\to q\bar{q}$
yields the production of a pair of quarks, the up-quark
for definitness) are shown for three particle masses:
$m_\chi = 10\,$GeV with $\Omega_{\chi}h^2 = 7.5\times 10^{-5}$ (long-dashed),
$m_\chi = 200\,$GeV with $\Omega_{\chi}h^2 = 1\times 10^{-4}$ (solid), and
$m_\chi = 4\,$TeV~\cite{remark6} 
with $\Omega_{\chi}h^2 = 6\times 10^{-4}$ (dashed-dotted).
It is evident that for decay times around $\tau\approx 10^3$s an efficient
destruction of \li7 is obtained.
For $\tau$ much shorter than $10^3$s the destroyed \be7 is regenerated, whereas
for $\tau$ much longer, incomplete \li7 burning in the reaction chain
\be7$(n,p)$\li7$(p,\alpha )$\he4 results in only partial reduction of the
total \li7 yield. As anticipated, the destruction of \li7 is accompanied by
production of D. When compared to the injection of thermal neutrons, D/H yields
are higher. This is due to D generated in the nuclear cascade itself (i.e. by
\he4 spallation and pionic fusion).
Cascade generated deuterium (as well as \h3, \he3, and \li6) is substantially
reduced per injected neutron for sources which inject nucleons with a soft
spectrum. For example, I have also employed a soft source with monoenergetic
nucleons of $250\,$MeV. Results for this case are shown by the short-dashed line,
assuming $\Omega_{\chi}h^2/m_{\chi} 
\approx 7.5\times 10^{-7}{\rm GeV^{-1}}$ 
and the injection
of one $np$ pair per decay~\cite{remark5}. A cascade $n$/D$\,\approx 10$ ratio
at $T\approx 40\,$keV is obtained in such scenarios. The more pronounced depth
of the \li7 dip in Fig. 2, by comparable (or smaller)
D production, indicates that
soft sources may affect a larger destruction factor in \li7 while respecting an
appropriate upper limit on the primordial D/H-ratio.
 
Concerning production of \li6 it is observed that yields of the order
of \li6/\li7$\, \approx 5\times 10^{-2}$, 
as observed in extreme low-metallicity
Pop II stars, are obtained for decay times in the range
$\tau\approx 1-2\times 10^3$s. 
For such $\tau$, substantial \li7 destruction is also observed, since
minimal \li7 production usually occurs for $\tau$ 
between 800 and $10^3$s.
The slight mismatch between the ``perfect'' decay time for \li7
destruction when compared to that for \li6 production is related to 
the more efficient destruction of \li6 (via \li6$(p,\alpha )$\he3) 
as compared to the analogous reaction for \li7. It 
also depends on the total \li6 yield per decaying particle and may
disappear if their exists a factor two underestimate in this 
quantity. However, it is not necessary that both decay times are
equal, as even when not, observationally acceptable scenarios may result.
This is illustrated in Table 1, which
shows the abundance yields of a few selected models, generally in
good agreement with the observational value of 
all three isotopes D, \li7, and \li6. 
If it all than the abundance of D $\sim 4\times 10^{-5}$
in these decaying particle scenarios is somewhat large.

It is tempting to adjust for this slight mismatch by resorting to somewhat
less in the literature
discussed possibilities of hadronic energy injection than that of a massive
decaying particle (such as the gravitino). 
These may include
unstable Q-balls~\cite{Enqv:03} 
or semistable strange-quark matter nuggets formed during a
QCD-transition~\cite{Witt:84}, 
among others. What is desired is a prolongation of the injection,
as compared to the decay of a particle. This may be, for example, due to
a population of Q-balls of varying sizes.
As a detailed analysis is beyond the scope of this
exploratory paper, I simply model the temporal injection of such a putative
source by $dn_{Q}/dt = n_{Q}^0/\tau\, {\rm min}[(t/\tau)^{-3},1]$, including
a long time injection tail. Here the spectrum of the primary and secondary
cascade products is taken to be that of a $200\,$GeV $q\bar{q}$ event and
$\Omega_{Q}h^2 = 10^{-4}$ has been assumed. Results for this case are given
by the dotted line in Fig.2, and for one particular $\tau$ on the last row
of Table 1.
It is seen, that for such "extended" emission
all, the D/H, \li7/H, and \li6/\li7 abundance constraints may be well
satisfied.
As another possibility I mention sources 
which have an unusually large ratio between electromagnetic
and hadronic energy injection ($\sim 10^2-10^3$). 
If emission lasts to around times of $10^5$s, a substantial fraction 
of D (as well as \li6) 
may be photodisintegrated. Nevertheless, such sources have
subsequently to stop radiating fairly abruptly, as otherwise an overproduction
of \li6 (either via \li7 photodisintegration or electromagnetic cascade
nucleosynthesis~\cite{Jeda:00}) results. Note, that this is also the reason
that the recently proposed \li7 photodisintegration as a solution to the comparatively
high predicted \li7 abundance may not apply~\cite{Feng:03}.

\begin{footnotesize}
\begin{table}
\newcommand{\lstrut}{{$\strut\atop\strut$}}
\caption{Light-element abundances for a few selected decaying particle
and extended emission scenarios. All models assumme $\Omega_bh^2= 0.026$.
See text for further details.}
\label{T:oh2}
\vspace{2mm}
\begin{center}
\begin{tabular}{|c||c|c|c|c|c|}
\hline
 model  & $\tau$(s) & D/H & \li7/H & \li6/\li7  \\
\hline
thermal neutrons       
& $7\cdot 10^2$ & $3.3\cdot 10^{-5}$ & $1.7\cdot 10^{-10}$ & $\simle 10^{-3}$ \\
soft source            
& $1.8\cdot 10^3$ & $4.0\cdot 10^{-5}$ & $1.9\cdot 10^{-10}$ & 0.03 \\
decay $m_{\chi} = 10\,$GeV  
& $1.5\cdot 10^3$ & $5.1\cdot 10^{-5}$ & $1.4\cdot 10^{-10}$ & 0.04 \\
decay $m_{\chi} = 200\,$GeV 
& $1.5\cdot 10^3$ & $4.1\cdot 10^{-5}$ & $2.1\cdot 10^{-10}$ & 0.03 \\
decay $m_{\chi} = 4\,$TeV  
& $1.5\cdot 10^3$ & $4.0\cdot 10^{-5}$ & $2.4\cdot 10^{-10}$ & 0.03 \\
extended emission 
& $7.1\cdot 10^2$ & $3.6\cdot 10^{-5}$ & $1.5\cdot 10^{-10}$ & 0.05 \\
\hline
\end{tabular}
\end{center}
\end{table}
\end{footnotesize}

Extended emission during BBN may result as well from annihilating particles.
Since a supersymmetric neutralino, in case it is the lightest
supersymmetric particle (LSP) and when R-parity is conserved, leads to
residual annihilation during BBN it is of interest to explore this case. 
In Fig. 3 I show the nucleosynthetic signatures of annihilating
neutralinos in D/H, \li7/H, and \li6/\li7 as a 
function of \Obh. Here the light solid, long-dashed, and dotted lines 
correspond to residual hadronic annihilation from light 
$5$ and $10\,$GeV neutralinos with varying s-wave annihilation cross sections
$\langle\sigma v\rangle =3$ and $5\times 10^{-26}{\rm cm^3/s}$. These values
of $\langle\sigma v\rangle$ correspond approximately to those required
for a thermal freeze-out of neutralino annihilations before (and after) the
QCD transition to yield $\Omega_{\chi}h^2\approx 0.1126$. Though such models
formally violate the LEP lower bound on the mass of the lightest neutralino
$m_{\chi}\simge 50\,$GeV, this latter bound implicitly
assumes gaugino mass unification at the GUT scale. When this assumption is
dropped, $m_{\chi}\approx 5\,$GeV neutralinos remain a viable dark matter
candidate~\cite{Bott:02}.
It is seen, that
with respect to SBBN, the \li7 overproduction problem is hardly
alleviated, while
such models produce \li6 in excess of that observed in Pop II stars. 
\li6 overproduction results particularly
for lower $m_{\chi}$ and larger $\langle\sigma v\rangle$, as here
residual annihilation is at it's strongest. 
To obtain the desired destruction of \li7 models would lead to
a factor $10-20$ overproduction of \li6. 
Models with p-wave annihilation $\langle\sigma v\rangle \sim T/m_{\chi}$
fare somewhat better, but at the expense of introducing exessively
large annihilation rates. 
In any case, residual annihilation of dark matter
particles seems not to be capable of explaining at the same time, 
the apparent SBBN \li7
overproduction problem and the \li6 abundance in low-metallicity stars.

\bef
\epsfxsize=8cm
\epsffile[100 75 300 415]{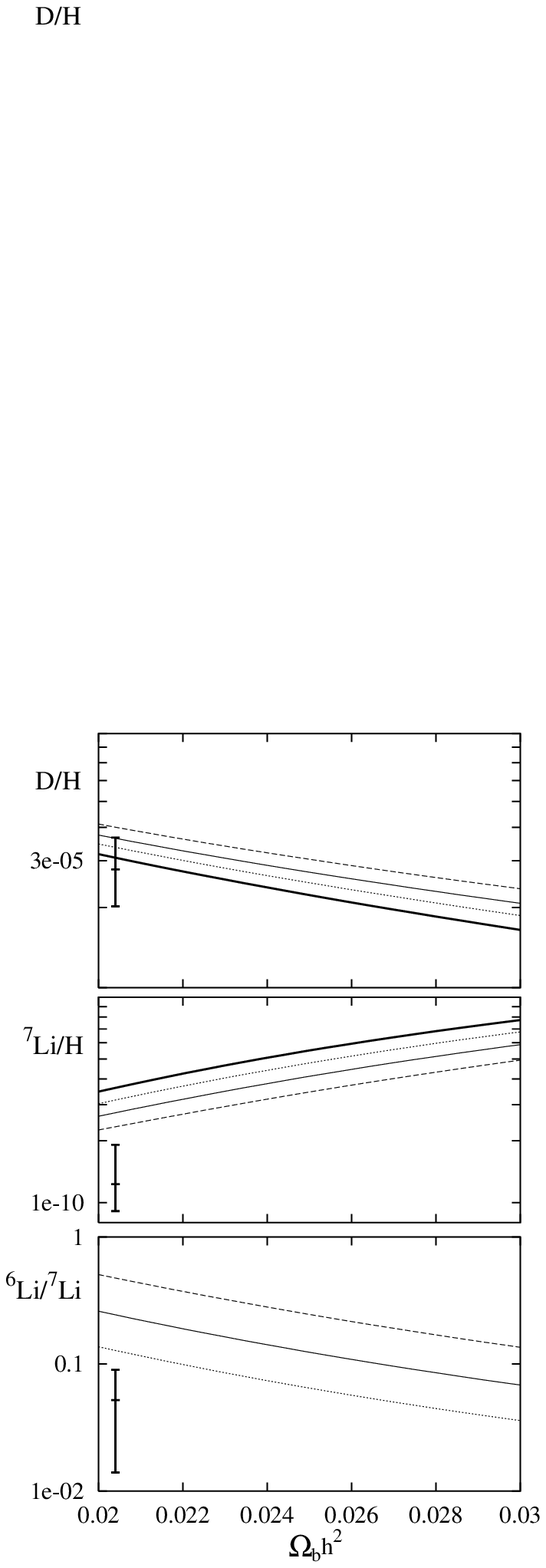}
\caption{Abundance yields of D/H, \li7/H, and \li7/\li6 in the
presence of residual neutralino annihilation for a neutralino
fractional contribution to the critical density of 
$\Omega{\chi}h^2 = 0.1126$. Three models are shown:
$m_{\chi} = 5\,$GeV and $\langle\sigma v\rangle = 3\times 10^{-26}{\rm cm^3/s}$
(light solid),
$m_{\chi} = 5\,$GeV and $\langle\sigma v\rangle = 5\times 10^{-26}{\rm cm^3/s}$
(dashed), and 
$m_{\chi} = 10\,$GeV and $\langle\sigma v\rangle = 3\times 10^{-26}{\rm cm^3/s}$
(dotted). For comparison, the SBBN yields are shown by the heavy solid lines.
Note the absence of appreciable \li6 production in SBBN.
Observational data on the D/H, \li7/H, and \li6/\li7 ratios indicated in
the figure are as in Fig.2.} 
\label{fig3}
\eef

I have so far implicitly assumed that ratios of D/H $\sim 3-4\times 10^{-5}$
are observationally allowed. This may seem somewhat 
at odds with the observational
determination of D/H$=2.78^{+0.44}_{-0.38}\times 10^{-5}$~\cite{Kirk:03}.
In this context, it should be noted (as Ref.~\cite{Kirk:03} did), that the 
formal errors given above may represent an underestimate of the true error
(due to further unknown systematics). 
As a spurious trend of D/H with total hydrogen column density of the
Lyman-$\alpha$ absorbers exists, with lower column density systems giving
higher inferred D/H (such as Q1009+2956 with 
$4.0\pm 0.65\times 10^{-5}$~\cite{Burl:98a} or
PKS 1937-1009 with $3.25\pm 0.3\times 10^{-5}$~\cite{Burl:98b}), 
I regard primordial values of
D/H $\sim 3-4\times 10^{-5}$ as viable. Similarly, a baryonic density
parameter of \Obh $= 0.026$ may seem compatible with the determination of
\Obh when WMAP data is only considered, but at odds with the estimate of
the combined data of WMAP and large-scale structure surveys (as well as 
Lyman-$\alpha$ forest surveys) which yields $0.0226\pm 0.0008$. Nevertheless,
since such analysis invokes assumptions of the underlying cosmological 
model, formal error bars should be taken with a grain of salt. Evidently, 
scenarios which are accompanied by additional late-time
photodisintegration of D (and \li6) may operate at lower \Obh.

It is interesting to speculate on which relic, in fact, could
cause the depletion of \li7 and concommitant production of \li6. 
Though residual annihilation of a dark matter particle usually predicts too
large of a \li6 abundance, when requiring a factor $2-3$ \li7 depletion, 
it may still represent the main source of \li6 in low-metallicity stars.
Alternatively, an initially high \li6 may (along with the \li7 )
be somewhat depleted by stellar processes to reach agreement with
observations on both isotopes. Other possibilities include the
evaporation of Q-balls or strange quark matter nuggets, but will not be
further discussed here. In light of the widely believed existence
of a supersymmetric (SUSY) extension of the standard model of particle physics,
particle decay may seem particularly promising, as there exist a variety
of possibilities involving the gravitino. 
When spontaneous SUSY breaking in a hidden sector is communicated to the
visible sector via gravitational interactions, or when breaking occurs
due to the super-Weyl anomaly, 
the gravitino is heavy
$\sim 0.1-100\,$TeV and long-lived. Moreover, it is easily produced
in the primordial plasma due to two-body scatterings at high temperture
$T_R$ with abundance 
$\Omega_{\tilde{G}}\sim 2\times 10^{-4}(T_R/10^8{\rm GeV})
({\rm TeV}/m_{\tilde{G}})$~\cite{Gabundance}.
It's decay into a gauge boson 
$B$ and it's superpartner $\tilde{B}$ occurs with life time
$\tau_{\tilde{G}}\approx 4\times 10^5\,{\rm s}N_c^{-1}
(m_{\tilde{G}}/1\, TeV)^{-3}$, where $N_c$ is the number of 
available decay channels
($N_c$ = 1 for the $U(1)_Y$ gauge boson and $N_c=8$ for gluons).
For decay into gluons one finds $\tau_{\tilde{G}}\approx 700\,$s for
a 4 TeV gravitino~\cite{remark20}, seeming somewhat large for the simplest
models of supergravity and somewhat small for models of anomaly-mediated
SUSY breaking. Other scenarios may result when the
gravitino itself is the LSP (lightest supersymmetric particle), such
as in gauge-mediated SUSY breaking (GMSB). In this case, the particle
next in the mass hierarchy (NLSP) undergoes thermal freeze-out of
annihilation reactions, with typical abundances of the order
$\Omega_{\chi}h^2\sim 10^{-4}-10^{4}$ independent of the reheat
temperature. The NLSP later on decays into the gravitino LSP and it's
superpartner with decay time $\tau_{\chi}\approx 240\,{\rm s}
(m_{\tilde{G}}/{\rm GeV})^2(m_{\chi}/{\rm 300\,GeV})^{-5}$, thus
making decay occur naturally at the desired $10^3$s for GeV gravitinos.
In GMSB scenarios the NLSP is frequently the bino $\tilde{B}$, 
right-handed stau $\tilde{\tau}$, and in a smaller part of the parameter
space the higgsino $\tilde{H}$. Typical annihilation rates for
$\tilde{B}$, $\tilde{\tau}$, and $\tilde{H}$ are in the ballpark
$10^{-27}$ for the former and $10^{-25}{\rm cm^3/s}$ for the latter
two particles, yielding thermal freeze-out
$\Omega_{\chi}$ of $\sim 1$ and $10^{-2}$, respectively. Nevertheless, a
significant spread around these reference values exists.
As I have shown
an $B_h\Omega_{\chi}h^2\approx 2\times 10^{-4}$ is preferable, where
$B_h$ is the hadronic branching ratio capable of producing nucleons
(i.e. excluding hadronic decay only generating pions as in the case of the
$\tau$). 
In the case of the $\tilde{\tau}$ and the $\tilde{H}$ the typical
abundance comes close to this value. 
The $\tilde{\tau}$, however, has only a small
phase-space supressed
$B_h\sim g_2^2/(32\pi^2)\sim 10^{-3}$ due to the coupling of $\tau$ to
$\nu_{\tau}$ and $W^{\pm}$, with the latter decaying into $q\bar{q}$
in seventy per cent of all cases. Here $g_2$ is the $SU(2)$ gauge
coupling constant.
An NLSP stau may thus only work
for $\Omega_{\tilde{\tau}}\sim 0.1-1$, demanding it to be rather heavy
$m_{\tilde{\tau}}\sim 1\,$TeV. For
a Higgsino one has $B_h\sim 0.1 - 1$ due to it's decay into heavy
quarks or massive gauge bosons, such that a Higgsino NLSP may have
the desired properties. Such scenarios may also be consistent with
the gravitino being the dark matter particle and successful
leptogenesis occurring at reheat temperatures 
$T_R\sim 10^{10}$GeV~\cite{Bolz}.
Finally, a typical bino abundance seems
too large since the bino $B_h$ is usually appreciable 
($\sim 0.1$) due to decay into the Z boson. Nevertheless, in this case
$B_h$ may be
phase space suppressed for $m_{\tilde{B}}$ not much larger 
than the Z mass. Particularly interesting in this case of $B_h\ll 1$ is
also the possible later photodisintegration of D, possibly removing the
requirement to operate at $\Omega_bh^2\simge 0.025$. Such decay
is also associated with an, albeit small, 
component of gravitino hot dark matter~\cite{JLM}. Last but not least,
the particle with the required properties may also be a super-WIMP
occurring in Kaluza-Klein theories~\cite{Feng}.

In summary, I have presented first results of a newly developed Monte Carlo
code examining the nuclear cascade development and cascade nucleosynthesis
resulting from the putative injection of energetic nucleons during the epoch
of BBN. Such an injection may result by a variety of means, such as, the decay of
gravitinos, the decay of NLSP's to gravitinos, 
the evaporation of Q-balls or strange quark matter nuggets, or the
residual annihilation of light neutralinos, among others. This most detailed 
numerical tool of this sort to date, has been applied in the regime of a 
"comparatively" weak perturbation during BBN leaving the primordial
\he4 abundance virtually unchanged when compared to SBBN. However, sources of
this kind may still have dramatic effects on the synthesized D, \li7, and \li6
yields. In particular, I have found that due to injection
of neutrons near $\tau\approx 10^3$s an efficient reduction of the final
\li7 yield may result. Such a factor 2-3 reduction of \li7 is just what is needed
to resolve the tension between the observationally inferred low
\li7/H ratios in low-metallicity Pop II stars and the theoretically predicted
high \li7/H ratio in SBBN. Of course, this holds true only if the
\li7 in these stars is not subject to some significant stellar depletion. 
In the simplest scenarios, destruction of \li7
is found to be accompanied by some production of D, though in
more complicated scenarios such additional D may later be partially 
photodisintegrated. If photodisintegration is absent, 
viable scenarios may only result
for $\Omega_bh^2\simge 0.025$, such that future (and current)
high precision determinations of \Obh by measurements of CMB anisotropies are
important for the evaluation of these models. Injection of energetic nucleons towards
the end of BBN may also result in the synthesis of the \li6 isotope, with
resultant \li6/\li7 isotope ratios comparable to those observed in low-metallicity
Pop II stars. This may present a viable alternative to the problematic explanation
of \li6 abundances in Pop II stars due to traditional cosmic ray nucleosynthesis..
It is intriguing to note that BBN with a weak non-thermal hadronic source, 
shows
good potential to resolve two current discrepancies in nuclear astrophysics.
In this context, particularly promising seems the decay of NLSP's to
LSP gravitinos of mass $\sim 1$GeV with NLSP freeze-out abundance 
either small $\Omega_{\chi}\approx 2\times 10^{-4}$ when an appreciable
hadronic (nucleonic) branching ratio $B_h$ exists, or larger when
$B_h\ll 1$. I have shown that this may the case for the NLSP being
either the bino, stau, or Higgsino.
Another viable possibility is the decay of $\simge {\rm TeV}$
gravitinos of abundance $\Omega_{\tilde{G}}\approx 5\times 10^{-4}$
produced during reheating at a reheat temperature of $T_R = 10^8$GeV.
I believe that such processes deserve further
consideration as, if indeed they occurred, they
may provide an invaluable source of information about the evolution of the
early Universe and the properties of a primordial relic.

\vskip 0.1in
I acknowledge helpful support and discussions with
E. Keihanen, G. Moultaka., J. Bystricky, and M. Chadeyeva.


\begin{thebibliography}{99}


%
\bibitem{reviews} R. A. Malaney and G. J. Mathews, Phys. Rept.
{\bf 229}, 145 (1993); S. Sarkar, Rept. Prog. Phys. {\bf 59},
1493 (1996)

\bibitem{Sper:03}
D.~N.~Spergel {\it et al.},
Astrophys.\ J.\ Suppl.\  {\bf 148}, 175 (2003)

\bibitem{Kirk:03}
D.~Kirkman, D.~Tytler, N.~Suzuki, J.~M.~O'Meara and D.~Lubin,
arXiv:astro-ph/0302006.

\bibitem{remark0} If not state otherwise, quoted error bars are
one-sigma.

\bibitem{remark1}
For example, for the same \Obh Ref.~\cite{Burl:00} 
obtain $4.67\times 10^{-10}$, to be compared to the 
lower $3.82\times 10^{-10}$
by Ref.~\cite{Cybu:03} 
and $4.15\times 10^{-10}$ by Ref.~\cite{Coc:04}
at even higher \Obh $\, = 0.0224$,
as well as $4.9\times 10^{-10}$ by Ref.~\cite{Cuoc:03} 
at \Obh$\, = 0.023$, indicating
approximate uncertainties in the predictions.

\bibitem{Burl:00}
S.~Burles, K.~M.~Nollett and M.~S.~Turner,
Astrophys.\ J.\  {\bf 552}, L1 (2001)

\bibitem{Cybu:03}
R.~H.~Cyburt, B.~D.~Fields and K.~A.~Olive,
astro-ph/0312629

\bibitem{Coc:04}
A.~Coc, E.~Vangioni-Flam, P.~Descouvemont, A.~Adahchour and C.~Angulo,
astro-ph/0401008

\bibitem{Cuoc:03}
A.~Cuoco, F.~Iocco, G.~Mangano, G.~Miele, O.~Pisanti and P.~D.~Serpico,
astro-ph/0307213

\bibitem{Ryan:99}
S.~G.~Ryan, T.~C.~Beers, K.~A.~Olive, B.~D.~Fields and J.~E.~Norris,
Astrophys. J. Lett. {\bf 530}, L57 (2000)

\bibitem{Boni:02}
P.~Bonifacio, et al., Astron. Astrophys. {\bf 390}, 91 (2002)

\bibitem{Aspl:03}
M.~Asplund, M.~Carlsson and A.~V.~Botnen,
Astron.\ Astrophys.\  {\bf 399}, L31 (2003)

\bibitem{Lemo:97}
M.~Lemoine, D.~N.~Schramm, J.~W.~Truran and C.~J.~Copi,
Astrophys. J. {\bf 478} 554 (1997)

\bibitem{Vang:99} 
E.~Vangioni-Flam, et al. New Astronomy {\bf 4} 245 (1999)

\bibitem{Sala:01}
M.~Salaris and A.~Weiss,
Astron. Astrophys. {\bf 376}, 955 (2001)

\bibitem{Thea:01}
S.~Theado and S.~Vauclair,
Astron. Astrophys. {\bf 375}, 70 (2001)

\bibitem{Pins:01}
M.~H.~Pinsonneault, G.~Steigman, T.~P.~Walker, .~K.~Narayanans and V.~K.~Narayanan,
Astrophys. J. {\bf 574}, 398 (2002)

\bibitem{Luri:03}
V.~Luridiana, A.~Peimbert, M.~Peimbert and M.~Cervino,
Astrophys.\ J.\  {\bf 592}, 846 (2003)

\bibitem{Izot:03}
Y.~I.~Izotov and T.~X.~Thuan,
astro-ph/0310421

\bibitem{Jeda:94}
K.~Jedamzik, G.~M.~Fuller and G.~J.~Mathews,
Astrophys.\ J.\  {\bf 423}, 50 (1994)

\bibitem{Reno:88}
M.~H.~Reno and D.~Seckel,
Phys.\ Rev.\ D {\bf 37}, 3441 (1988)

\bibitem{Kohr:01}
K.~Kohri,
Phys.\ Rev.\ D {\bf 64}, 043515 (2001)

\bibitem{Dimo:88}
S.~Dimopoulos, R.~Esmailzadeh, L.~J.~Hall and G.~D.~Starkman,
Astrophys.\ J.\  {\bf 330}, 545 (1988)

\bibitem{Khlop}
Y.~L.~Levitan, I.~M.~Sobol, M.~Y.~Khlopov and V.~M.~Chechetkin,
Sov.\ J.\ Nucl.\ Phys.\  {\bf 47}, 109 (1988); 
E.~V.~Sedelnikov, S.~S.~Filippov and M.~Y.~Khlopov,
Phys.\ Atom.\ Nucl.\  {\bf 58}, 235 (1995).

\bibitem{li6:highZ}
V.~V.~Smith, D.~L.~Lambert, and P.~E.~Nissen,
Astrophys.~J. {\bf 408} 262 (1993); {\bf 506} 405 (1998);
L.~M.~Hobbs and J.~A.~Thorburn,
Astrophys.~J. {\bf 491} 772 (1997); R.~Cayrel, M.~Spite, F.~Spite,
E.Vangioni-Flam, M.~Cass\'e, and J.~Audouze, Astron. \& Astrophys. {\bf 343}
923 (1999); P.~E.~Nissen, M.~Asplund, V.~Hill, and S.~D'Odorico,
Astr.~\&~ Astrophys. {\bf 357} L49 (2000)

\bibitem{li6:lowZ} P.~E.~Nissen, D.~L.~Lambert, F.~Primas, and V.~V.~Smith,
Astr.~\&~ Astrophys. {\bf 348} 211 (1999)

\bibitem{remark2} It may be that this is simply a selection
effect as most of these claimed detections are only around 2$\sigma$.

\bibitem{Noll:97} K.~M.~Nollett, M.~Lemoine, and D.~N.~Schramm,
Phys.~Rev. {\bf C56} 1144 (1997)

\bibitem{pro:li6}
E.~Vangioni-Flam, M.~Cass\'e, R.~Cayrel, J.~Audouze,
M.~Spite, and F.~Spite, New~Astron. {\bf 4} 245 (1999);
B.~D.~Fields and K.~A.~Olive, New~Astron. {\bf 4} 255 (1999)

\bibitem{Rama:99}
R.~Ramaty, S.~T.~Scully, R.~E.~Lingenfelter and B.~Kozlovsky,
Astrophys. J. {\bf 534}, 747 (2000)

\bibitem{Alib:02}
A.~Alib\'es, J.~Labay, and R.~Canal,
Astrophys. J. {\bf 571}, 326 (2002)


\bibitem{Suzu:02}
T.~K.~Suzuki and S.~Inoue,
Astrophys. J. {\bf 573} 168 (2002)

\bibitem{Jeda:00}
K.~Jedamzik,
Phys.\ Rev.\ Lett.\  {\bf 84}, 3248 (2000)

\bibitem{remark3} Such considerations are also very useful to derive
constraints on late-decaying particles in the early Universe.

\bibitem{remark4} The analysis does currently not include the effects from
antinucleons. An antinucleon scatters not more than 1-2 times on nucleons
before an annihilation reaction occurs. Due to the small \he4/H ratio it thus
has a probability of roughly $10\%$ to annihilate on \he4, thereby producing
around $\sim 0.3$ D, \he3, and \h3 per annihilation. One may therefore
expect $\sim 0.03$ D, (\he3 and \h3) per injected antinucleon. This should
be compared to $\sim 1$ D (see text further below) produced per annihilation
by injected energetic nucleons, since those latter may scatter of the order
10-20 times before falling below the threshold for \he4 spallation.

\bibitem{Smit:93}
M.~S.~Smith, L.~H.~Kawano and R.~A.~Malaney,
Astrophys.\ J.\ Suppl.\  {\bf 85}, 219 (1993)

\bibitem{Angu:99} C.~Angulo, M.~Arnould, M.~Rayet, et al., Nucle Phys. {\bf A656} 3 (1999)

\bibitem{pythia}T. Sj\"ostrand, P. Ed\'en, C. Friberg, L. L\"onnblad, G. Miu, S. Mrenna and E. Norrbin, 
Computer Physics Commun. {\bf 135} 238 (2001);  
www.thep.lu.se/~torbjorn/Pythia.html

\bibitem{remark7}
For higher particle masses $m\simge 1\,$TeV, the ratio of 
secondary to primary
neutrons is more of the order of $\sim 10$.

\bibitem{remark6} Note here, that predictions become less accurate
for higher mass particles due to poor reaction rate data at
high energies. Whereas the typical energy of a nucleon 
generated during the decay of a
$m_{\chi} = 200\,$GeV particle is 8 GeV , it is 55 GeV for a
$m_{\chi} = 4\,$TeV particle.

\bibitem{remark5}
The decay may still be envisioned as baryon-number non-violating, as it may be
associated with the injection of one $\bar{n}\bar{p}$ pair. 
These latter particles
predominantly annihilate on protons, thus effectively
yielding the injection of one energetic np pair, and the annihilation of two 
thermal protons. In any case, due to the weakness of the source, the removal,
or non-removal, of thermal protons has a negligible effect on the 
final result. 
As argued above~\cite{remark4}, 
cascade produced D due to antinucleon-annihilation on \he4 is a 
subdominant 
$\sim 10\%$ effect, which nevertheless, becomes stronger for lower
energy primary nucleons.

\bibitem{Enqv:03}
K.~Enqvist and A.~Mazumdar,
Phys.\ Rept.\  {\bf 380}, 99 (2003) 

\bibitem{Witt:84}
E.~Witten,
Phys.\ Rev.\ D {\bf 30}, 272 (1984)

\bibitem{Feng:03}
J.~L.~Feng, A.~Rajaraman and F.~Takayama,
Phys.\ Rev.\ D {\bf 68}, 063504 (2003)

\bibitem{Bott:02}
A.~Bottino, N.~Fornengo and S.~Scopel,
Phys.\ Rev.\ D {\bf 67}, 063519 (2003)

\bibitem{Burl:98a}
S.~Burles and D.~Tytler,
arXiv:astro-ph/9712109.

\bibitem{Burl:98b}
S.~Burles and D.~Tytler,
Astrophys.\ J.\  {\bf 499}, 699 (1998)

\bibitem{remark20} 
Even when the decay channel into gluons, for example, due to
kinematic reasons is disallowed, decay into the $U(1)_Y$ B-gauge boson
may induce the injection of nucleons, since the B is a mixture
between the photon and the Z, with the latter having an appreciable
branching ratio into quarks. Nevertheless, in this case a gravitino mass
of the order of 8 TeV is required.

\bibitem{Gabundance}
T.~Moroi, H.~Murayama and M.~Yamaguchi,
Phys.\ Lett.\ B {\bf 303}, 289 (1993);
M.~Bolz, A.~Brandenburg and W.~Buchmuller,
Nucl.\ Phys.\ B {\bf 606}, 518 (2001)

\bibitem{Bolz}
M.~Bolz, W.~Buchmuller and M.~Plumacher,
Phys.\ Lett.\ B {\bf 443}, 209 (1998)

\bibitem{JLM}
K.~Jedamzik, M.~Lemoine and G.~Moultaka,
in preparation

\bibitem{Feng}
J.~L.~Feng, A.~Rajaraman and F.~Takayama,
Phys.\ Rev.\ Lett.\  {\bf 91}, 011302 (2003)

\end{thebibliography}
\end{document}